\documentclass[%
 aip,
 amsmath,amssymb,
 reprint,%
]{revtex4-1}

\usepackage{graphicx}
\usepackage{dcolumn}
\usepackage{bm}

\usepackage[utf8]{inputenc}
\usepackage[T1]{fontenc}
\usepackage{mathptmx}
\usepackage{etoolbox}
\usepackage{xcolor}

\makeatletter
\def\@email#1#2{%
 \endgroup
 \patchcmd{\titleblock@produce}
  {\frontmatter@RRAPformat}
  {\frontmatter@RRAPformat{\produce@RRAP{*#1\href{mailto:#2}{#2}}}\frontmatter@RRAPformat}
  {}{}
}%
\makeatother
\begin{document}


\title[LB-NGRC]{Locality Blended Next Generation Reservoir Computing For Attention Accuracy}
\author{Daniel J. Gauthier}
 \email{daniel.gauthier@flyrescon.com}
\affiliation{ResCon Technologies, LLC, 1275 Kinnear Rd., Suite 239, Columbus, Ohio 43212 USA}
\author{Andrew Pomerance}
\affiliation{ResCon Technologies, LLC, 1275 Kinnear Rd., Suite 239, Columbus, Ohio 43212 USA}
\affiliation{Potomac Research LLC, 801 N Pitt St., Suite 117, Alexandria, Virginia 22314 USA}

\author{Erik Bollt}%
\affiliation{Clarkson University, Department of Electrical and Computer Engineering, Potsdam, NY 13669, USA \\ Clarkson Center for
Complex Systems Science (C3S2), Potsdam, NY 13699, USA
}%

\date{\today}

\begin{abstract}
We extend an advanced variation of a machine learning algorithm, next-generation reservoir Computing (NGRC), to forecast the dynamics of the Ikeda map of a chaotic laser.  The machine learning model is created by observing time-series data generated by the Ikeda map, and the trained model is used to forecast the behavior without any input from the map.  The Ikeda map is a particularly challenging problem to learn because of the complicated map functions.  We overcome the challenge by a novel improvement of the NGRC concept by emphasizing simpler polynomial models localized to well-designed regions of phase space and then blending these models between regions, a method that we call locality blended next-generation reservoir computing (LB-NGRC).  This approach allows for better performance with relatively smaller data sets, and gives a new level of interpretability. We achieve forecasting horizons exceeding five Lyapunov times, and we demonstrate that the `climate' of the model is learned over long times.
\end{abstract}

\maketitle

\begin{quotation}
There is great interest in learning dynamical system models using only observed time series data.  Machine learning is an emerging tool for this purpose, but it often provides a `black box' solution that is hard to interpret.  Next-generation reservoir computing provides a more interpretable solution because it is based on a linear superposition of nonlinear functions, where the linear weights are learned during model training. It also has the advantage of requiring smaller training datasets, and reduces the computer time required to perform training. However, it can struggle with systems with complicated functions describing the dynamics.  Here, we show a major improvement, called locality blended next-generation reservoir computing (LB-NGRC), which can be extended by having it focus its attention on smaller regions of phase space, greatly improving its performance on a challenging problem.
\end{quotation}

\section{Introduction}\label{sec:Intro}

A current topic of dynamical systems research is using model-free, data-driven methods to learn the underlying system.  For chaotic dynamical systems, this involves nonlinear system identification using, for example, machine learning (ML) algorithms such as deep learning.\cite{Goodfellow2016}  The goal of the ML model is to forecast the future dynamics of the system or to predict variables that are inaccessible during model deployment.

One example of an ML model that excels in predicting chaotic dynamics is known as reservoir computing, which is an artificial neural network comprising a core `reservoir' of time-dependent nonlinear neurons connected in a recurrent topology.\cite{Jaeger2004,Maass2002}  Reservoir computing has been successfully applied to a variety of nonlinear dynamical systems and produces state-of-the-art results\cite{Pathak2017} but with vastly smaller training datasets and smaller compute times.  Its performance is comparable to that obtained with deep learning algorithms, as demonstrated by head-to-head comparisons.\cite{Vlachas2020,Bompas2020} Also, reservoir computing can be mapped onto physical systems for efficient analog realizations.\cite{Nakajima2021}

Next-generation reservoir computing (NGRC) simplifies the reservoir computing algorithm even further by separating aspects of the model into different compartments,\cite{NGRC} as described below. Here, time-delayed copies of the time series data generated by the dynamical system are passed through nonlinear functions, which are weighted by the model parameters and summed.  The NGRC algorithm has been successfully applied to various problems with greatly reduced training dataset requirements and compute times while providing state-of-the-art model accuracies.

In the original NGRC report, the nonlinear functions of the model are low-order monomials, motivated by the polynomial form of the equations describing the dynamical systems.  Suppose the modeler has no information about the underlying dynamical system, or it is known that polynomial functions do not describe it. In that case, they can be replaced by generic sigmoidal-shaped functions in a single-layer feedforward neural network with internal trainable weights\cite{Zhai2023} or with randomly assigned internal parameters in an extreme learning machine.\cite{ELM}  However, these approaches lose the interpretability of monomial-based features of the NRGC.

The primary purpose of this paper is to use an algorithm that retains the interpretability of monomial-based nonlinear functions in NGRC while modeling dynamical systems that are inherently not described by polynomials.  We advance a method known as piecewise-polynomial regression trees \cite{Chaudhuri1994} for this problem and select modern tools to simplify the task, particularly regarding the critical stage of developing a relevant partition.

\begin{figure*}[tb] 
    \centering
    \includegraphics[width=0.8\linewidth]{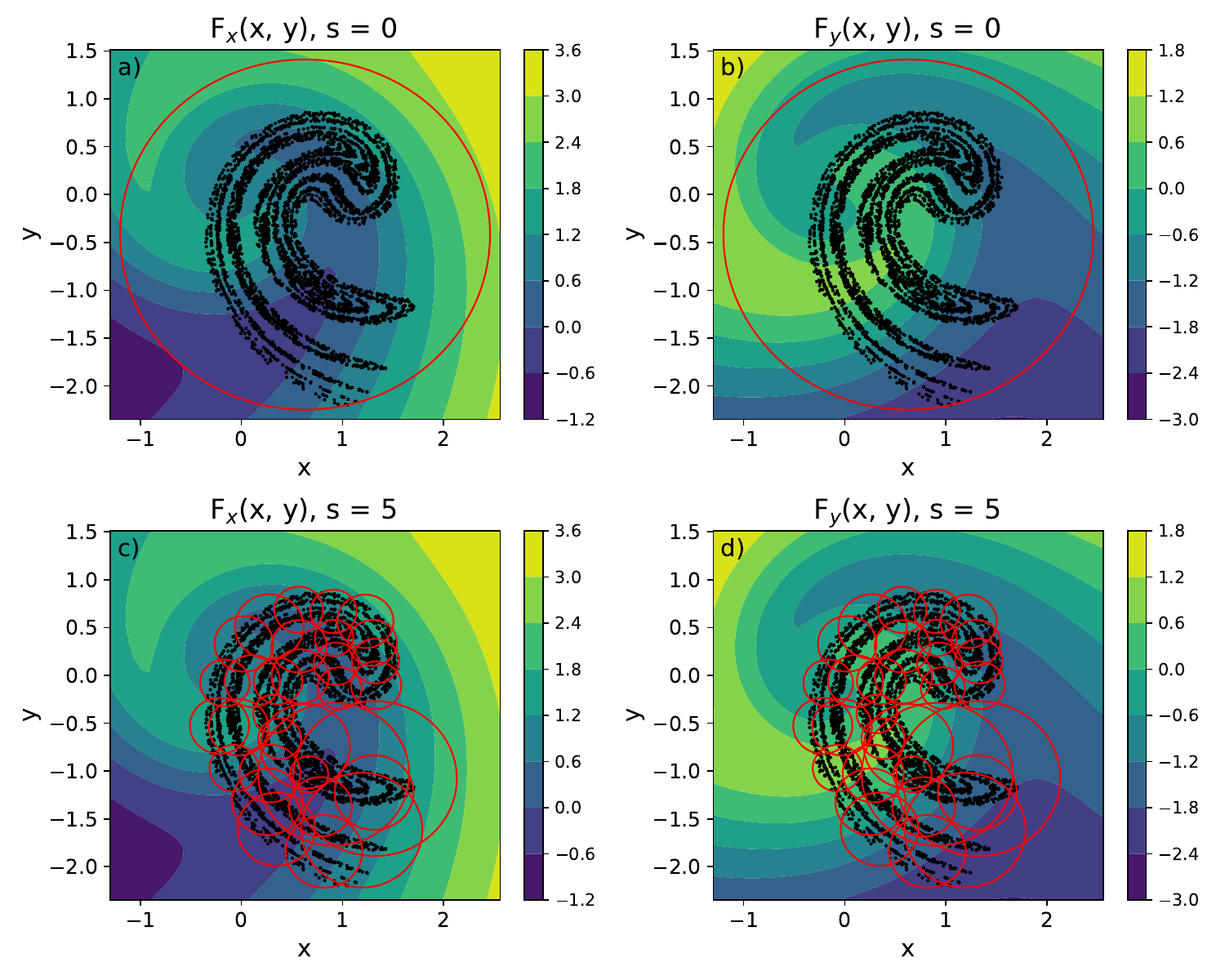}
    \caption{\textbf{The Ikeda attractor covered with balls.} Contour plots of the map functions superimposed with 4,000 points of the Ikeda attractor and the balls identified by the ball tree clustering algorithm for a scale of 0 (top row) and 5 (bottom row).
    }
    \label{fig:storyboard}
\end{figure*}

Figure~\ref{fig:storyboard} shows a dynamical system whose underlying functions have a swirling characteristic, which would be difficult to describe by a low-order polynomial. The key insight is adding an \textit{attention mechanism} to the model using a hierarchical tree structure to cluster the data. Here, a low-order monomial-based NGRC is used to describe the dynamics in a \textbf{local} neighborhood of phase space, as illustrated by the red circles in Fig.~\ref{fig:storyboard}, and models for all regions are smoothly \textbf{blended}. We refer to this model as locality blended NGRC, or LB-NGRC for short.

Our approach brings together multiple concepts and tools from the statistics and ML communities but has not been used in combination as put forward here to the best of our knowledge.  We demonstrate our approach using data generated by a low-dimensional chaos model known to be hard\cite{Lai2021} to model using other ML tools:  the two-dimensional Ikeda map\cite{Ikeda1980} describing the behavior of a chaotic laser.  We only consider the forecasting problem, that is, predicting the system state one step in the future, given its current state.

Because we are describing a new approach, we do not exhaustively apply the method to a wide range of dynamical systems.  Rather, we attempt to keep the description as simple as possible to encourage others to adapt this approach to their favorite problem. In the next section, we give background on NGRC and then introduce the attention mechanism in Sec.~\ref{sec:NGRCa}.  We present our results in Sec.~\ref{sec:Results} and follow with discussions in the last section.

\section{Background}\label{sec:Back}

In ML approaches, the algorithm is typically presented with discretely sampled time series data.  The data are divided into a part used for `training' the algorithm, that is, performing the system identification, and a `testing' part, which is used to validate the generalizability of the model by using it to predict data not seen during training.  A wide variety of ML algorithms excel at learning dynamical systems from time-series data, such as low-dimensional chaos described by maps or differential equations or complex spatial-temporal behavior described by partial differential equations.

Deep learning\cite{Goodfellow2016} and its variations provide state-of-the-art nonlinear system identification but often require large training datasets and long compute times on advanced hardware such as graphical processor units.  One reason for the long compute times is that the deep learning model parameters are embedded within nonlinear functions, which require computationally difficult nonlinear optimization.

An alternative to deep learning is to use ML models that are linear in the parameters to be learned.\cite{Sarangapani2006}  Here, the ML model is a linear superposition of nonlinear functions of the input data.  We stress that this does not require that the dynamical system is of this form; we only require the existence of a linear-in-the-parameters (LP) model that is a good approximation of the dynamical system. Finding the parameters only requires linear optimization methods, such as regularized least squares regression. As with deep learning models, it has been proven that LP models are universal function approximators when the models are large, so either approach can be applied to our problem.

We consider a dynamical system measured at equal discrete time steps and described by a mapping
\begin{equation}
\mathbf{x}_{n+1} = \Phi(\mathbf{x}_n,\mathbf{x}_{n-1},\ldots) \label{eq:flow}
\end{equation}
where $\mathbf{x}_n$ is the state of the system at time step $n$ and $\Phi$ is the dynamical system \textit{flow}.  The goal of NGRC is to learn an LP model of $\Phi$.

To this end, we define an LP model given by
\begin{equation}
\mathbf{x}_{n+1} = \mathcal{O}(\mathbf{x}_n,\mathbf{x}_{n-1},\ldots)\mathbf{W} \simeq \Phi(\mathbf{x}_n,\mathbf{x}_{n-1},\ldots), \label{eq:NGRC}
\end{equation}
where $\mathcal{O}$ is a so-called feature vector that contains nonlinear functions of the input data, and the elements of the matrix $\mathbf{W}$ must be `learned' or `trained' based on data generated by the dynamical system.  As an aside, Eq.~\ref{eq:NGRC} is written in a form typically used by the wider ML community to take advantage of existing toolsets.  The reservoir computing community often writes it as $\mathbf{W}\mathcal{O}$, which works equally well.  This change in convention requires rearranging Eq.~\ref{eq:ridge} below so our results may appear different than the expression found in other reservoir computing papers.   

We focus on the Ikeda map \cite{Ikeda1980} given by
\begin{eqnarray}
    x_{n+1} &=& F_x(x_n,y_n) = 1+u(x_n \cos{\theta_n}-y_n\sin\theta_n), \label{eq:Ikeda-x} \\
    y_{n+1} &=& F_y(x_n,y_n) = 1+u(x_n \sin{\theta_n}+y_n\cos\theta_n), \label{eq:Ikeda-y} \\
    \theta_n &=& 0.4 - \frac{6}{1+x_n^2+y_n^2}, \label{eq:Ikeda-phi}
\end{eqnarray}
with $u=0.9$.  Figure~\ref{fig:storyboard} shows iterations of the map superimposed on a contour plot of the functions $F_x(x_n,y_n)$ and $F_y(x_n,y_n)$.  The range of $x_n$ and $y_n$ is large enough so that the transcendental expressions in Eqs.~\ref{eq:Ikeda-x} and \ref{eq:Ikeda-y} saturation-like term in Eq.~\ref{eq:Ikeda-phi} are not well approximated by a Taylor series expansion; that is, as a function of monomials.  The difficulty in using an ML model to forecast the Ikeda dynamics has been pointed out previously.

We use knowledge of the Ikeda map to simplify the NGRC model: we assume that the feature vector only has nonlinear functionals of the current state of the system $\mathbf{x}_n = [x_n,y_n]$ (a $(1\times2)$ row vector) and no data from past states.  Then the $x$-component ($y$-component) of $\mathcal{O}\mathbf{W}$ is an LP approximation of $F_x(x_n,y_n)$ ($F_y(x_n,y_n)$).  That is, the ML algorithm learns a nonlinear function using only time series data generated by the mapping.

Motivated by the original paper on NGRC, the feature vector contains monomials up to order $N$ as
\begin{equation}
\mathcal{O}_n = [c,x_n,y_n,x_n^2,x_n y_n, y_n^2, \ldots,x_n^N,\ldots,y_n^N] \label{eq:poly_featvec},
\end{equation}
where $c$ is a constant. The problem with a monomial-based model is that the number of features grows exponentially with $N$. On the other hand, our advice is to try a low-order monomial-based model first because it is simple to evaluate and works well for a large number of problems.

Previous work \cite{Giona1991} finds an acceptable monomial-based model for the Ikeda map for $u=0.7$ and $N=5$.  Here, the dimension of $\mathcal{O}$ is $(1\times 21)$ and $\mathbf{W}$ has dimension $(21\times 2)$.  As shown below, the 5$^{th}$-order monomial model is not accurate for the larger value of $u$ used here.

Finding the LP model requires two steps.  The first step, or the training phase, assumes that we have access to the dataset one step in the future. This is known as supervised learning.  The second step is the testing or deployment phase, where only an initial value of $\mathbf{x}_i$ is known and the model is used to predict the states into the future by feedback of the output of the model to the input.  This is known as a generative model.

In the training phase, we form a sequence of feature vectors\cite{polynomial} using $M+1$ consecutive observations of $\mathbf{x}_n$ and vertically stack them in a matrix
\begin{equation}
\mathbf{O} = \mathcal{O}_{1} \oplus \mathcal{O}_{2} \oplus \ldots \oplus \mathcal{O}_{M}
\end{equation}
of dimension $(M\times N_{feat})$, where $\oplus$ indicates vertical concatenation.  Similarly, we form the `true' values we are learning, which is given by 
\begin{equation}
\mathbf{Y} = \mathbf{x}_2 \oplus \mathbf{x}_{3} \oplus \ldots \mathbf{x}_{M+1}
\end{equation}
for the one-step ahead forecasting problem we consider here.

The learned weights are found using regularized least square regression (ridge regression)\cite{Ridge} through the relation
\begin{equation}
\mathbf{W} = (\mathbf{O}^T\mathbf{O}+\alpha \mathbf{I})^{-1}\mathbf{O}^T\mathbf{Y},\label{eq:ridge}
\end{equation}
where $T$ is the transpose, $\mathbf{I}$ is the identity matrix, and $\alpha$ is the regularization parameter.  The procedure goes over to standard least squares regression when $\alpha=0$.  Non-zero values of $\alpha$ reduce errors in the numerical computation of the matrix inverse when the condition number of $\mathbf{O}^T\mathbf{O}$ is large (often the case in our application), and it also penalizes large components of $\mathbf{W}$.

\section{LB-NGRC is NGRC with attention}\label{sec:NGRCa}

The key concept of LB-NGRC is to add an attention mechanism to the LP model.  Here, we translate the coordinates to
\begin{equation}
\tilde{\mathbf{x}}_b = \mathbf{x} - \mathbf{x}_b
\end{equation}
and fit a polynomial model to the data in a small neighborhood of $\mathbf{x}_b$ with corresponding feature vectors $\mathcal{O}_{n,b}$ and model parameters $\mathbf{W}_b$.  

The individual models are blended to give an overall LP model given by
\begin{equation}\label{eq:blendedNGRC}
\Phi(\mathbf{x}_n) \simeq \frac{\sum_{b=1}^B\mathcal{O}_{n,b}\mathbf{W}_b~\textrm{RBF}_b(\tilde{\mathbf{x}}_b,\sigma_b)}{\sum_{b=1}^B \textrm{RBF}_b(\tilde{\mathbf{x}}_b,\sigma_b)},
\end{equation}
where $\textrm{RBF}_b(\tilde{\mathbf{x}}_b,\sigma_b)$ is a radial basis function centered on $\mathbf{x}_b$ and a scalar parameter $\sigma_b$ that sets the spatial scale of the function.  There are many choices for the radial basis function; we find that a Gaussian function
\begin{equation}\label{eq:bend2}
\textrm{RBF}_b(\tilde{\mathbf{x}}_b,\sigma_b) = \mathrm{exp}\left( -\tilde{\mathbf{x}}_b^2/(2\sigma_b^2)\right)
\end{equation}
works well for modeling the Ikeda map.

One important aspect of our approach is that the individual monomial-based models $\mathcal{O}_{n,b}\mathbf{W}_b$ are smooth and defined over the entire phase space.\cite{Chaudhuri1994}  Joining the models requires selecting an RBF that focuses attention on the local neighborhood of phase space.  Typically, the spatial scale of the radial basis function, set by $\sigma_b$ is set to the typical spacing between neighborhood centers.  Importantly, we do not explicitly match local models at the boundaries of the neighborhoods as is often done for interpolation methods; the radial basis functions do this automatically.  A detailed analysis of the quality of fit is given in the Appendix \ref{app:qual}. 

So far, we have not addressed how to partition phase space into neighborhoods.  There is a wide variety of methods available from the ML community known as clustering algorithms that can partition the space.  These are typically unsupervised methods that break up the space using only the data.  Here, we use the \textit{ball tree} method, which uses a branching tree method to subdivide the data.

The ball tree method places a hypersphere (a ball) about clusters of neighboring data points.  On the coarsest scale (top row, Fig.~\ref{fig:storyboard}), it puts a single ball around the entire data set, where the center of the ball $\mathbf{x}_b$ is the average of the data along each axis and the ball radius $r_b$ is just large enough to encompass the data.  At the next scale, it divides the data in half and covers it with two balls, known as the children.  It selects the ball centers by maximizing the distance (using a user-defined metric) between the ball centers while minimizing the ball radii. The balls overlap, but each point is associated with a single ball. This procedure is applied recursively until the finest (leaf) level set by the user is attained or the data runs out, whichever happens first. For scale level $s$, there are $2^s$ balls, with an average of $M/2^s$ points per ball.  The bottom row of Fig.~\ref{fig:storyboard} shows the balls for $s=5$.

Once the ball tree is formed, it can be used to identify the neighbors of a new point using a fast search algorithm.  Importantly, the ball-tree method is efficient for high-dimensional data, so our method can be used for complex problems residing in a high-dimension phase space. Also, the hypersphere geometry is well matched to radial basis functions.

The cost of LB-NGRC is an increase in the model metaparameters that must be optimized.  For the standard NGRC model, the metaparamters include the number of training data points $M$, the maximum polynomial degree $N$, and the ridge regression regularization parameter $\alpha$.  Because we are working with a known mapping, the maximum delay-time and skip parameters for the standard NGRC are not needed here. The LB-NGRC model adds four additional metaparameters: the ball tree scale $s$, the choice of the radial basis function, the radius of the radial basis function relative to the ball radius $\sigma_b/r_b$, and the metric used for measuring distance. 

\section{Results}\label{sec:Results}

We apply LB-NGRC to the Ikeda map for 50,000 iterations with an initial condition on the attractor. The first 10,000 points are used for training, and the remainder is used for testing.  During testing, we break the testing data into 10 equal-size segments (4,000 points each), and we restart testing at the beginning of each to determine the variance in the testing error.  We find that it is important to test over such long intervals to ensure that the LB-NGRC model reproduces the `climate' of the attractor (model overfitting can lead to a collapse or divergence of the dynamics).

The first step is to use a ball tree clustering algorithm\cite{BallTree} that goes to a maximum scale of $s=6$ ($2^6=64$ balls). We use the default Minkowski metric, which is the Euclidean distance for the two-dimensional problem considered here). Figure~\ref{fig:storyboard} shows the ball coverage for $s=5$.   
We fit a low-order polynomial model\cite{polynomial, Ridge} to the data in each ball and determine the normalized root-mean-square error (NRMSE) using the training data \textit{without} blending the models. Here, the error is normalized to the standard deviation of the Ikeda attractor using an $L_2$ norm.  The training NRMSE assesses the quality of the fit of the model to the data.  The regularization parameter is optimized using a grid search to best reproduce the Ikeda attractor, as discussed below.

\begin{figure}[htb!] 
    \centering
    \includegraphics[width=0.9\columnwidth]{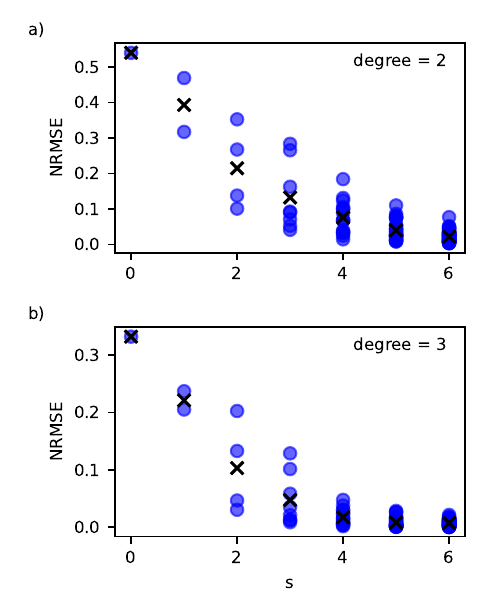}
    \caption{\textbf{Training error.} Error as a function of the ball tree scale for a locally a) quadratic model with $\alpha=1\times 10^{-4}$ and b) cubic model with $\alpha=1\times 10^{-5}$.  The black $\times$ indicates the mean of the errors for each ball.  
    }
    \label{fig:train_error}
\end{figure}

Figure~\ref{fig:train_error} shows how the fit changes with the ball scale for locally quadratic and cubic polynomials. Note that the average number of points in each ball decreases by a factor of two for each increase in scale because we are using a fixed number of training points.  Each blue symbol is the error for an individual ball, and the mean of the fits is also indicated.  The error and standard deviations of the fits decrease smoothly as the scale increases, where the cubic term has a somewhat lower error. If we turn off regularization ($\alpha=0$) so that the fitting is least squares regression, the errors at the highest scale are smaller, but this comes at the cost of greater testing error.

To assess the generalizability of the models, we test them at 10 different starting points on data it has not seen during training.  Because the system is chaotic, small errors will necessarily grow on a time scale of the Lyapunov time, which is the inverse of the positive Lyapunov exponents quantifying the rate of exponential divergence of the iterates in phase space.  Here, the Lyapunov time is 1.99,\cite{Brown1991} corresponding to about two iterations of the map. It is common practice to test over a single Lyapunov time, but our models are so accurate that we test over 5 Lyapunov times.  Over this interval, we find the NRMSE between the actual and predicted iterates.  Each testing segment corresponds to $\sim$2,000 Lyapunov times, so the segments are uncorrelated.

\begin{figure}[htb!] 
    \centering
    \includegraphics[width=0.9\columnwidth]{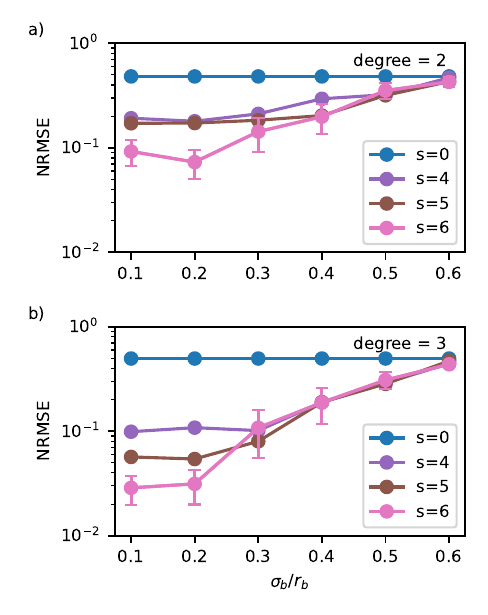}
    \caption{\textbf{Test error.} Mean testing NRMSE with different scales as a function of the radius of the radial basis function for 10 testing segments.  The error bars represent the error of the mean and are only shown for the finest scale for clarity.  The error bars are similar for the other scales. Regularization: locally a) quadratic model with $\alpha=1\times 10^{-4}$ and b) cubic model with $\alpha=1\times 10^{-5}$.
    }
    \label{fig:test_error}
\end{figure}

Figure~\ref{fig:test_error} shows the testing error as a function of the size of the radial basis function $\sigma$ relative to the ball radius $r_b$. We include the case $s=0$, which is the standard NGRC model and does not depend on the radial basis function radius.  We skip the smaller scales because the error tends to be large; we only begin to display the data for $s \geq 4$.

For reference, Giona \textit{et al.} find that a 5$^{th}$-order polynomial model (standard NGRC, $s=0$) accurately predicts the Ikeda attractor for a smaller bifurcation parameter $u$ than we use here.  For our value of $u$, we find that the testing NRMSE $\sim$ 0.5, which is comparable to random guessing.

We see that the finer balls (larger $s$) favor tighter attention (narrower radial basis functions) as the $s$ increases.  However, the NRMSE is not overly sensitive to this ratio.  For locally quadratic (cubic) models, we take $\sigma_b/r_b = 0.3$ (0.2) in the analysis below.  It is also evident that the error approximately saturates and does not decrease substantially for smaller $\sigma_b$.  The location of the breaking point depends on the ridge regression parameter, which requires some interactive optimization of the metaparameters.

A similar behavior occurs for the NRMSE as a function of the scale for fixed $\sigma_b$.  The error approximately saturates at a modest scale.  Again, the saturation scale value depends on the regression parameter.  Below, we take $s=5$, which trades off the accuracy of the LB-NGRC model and the number of model parameters.

\begin{figure*}[htb!] 
    \centering
    \includegraphics[width=0.8\linewidth]{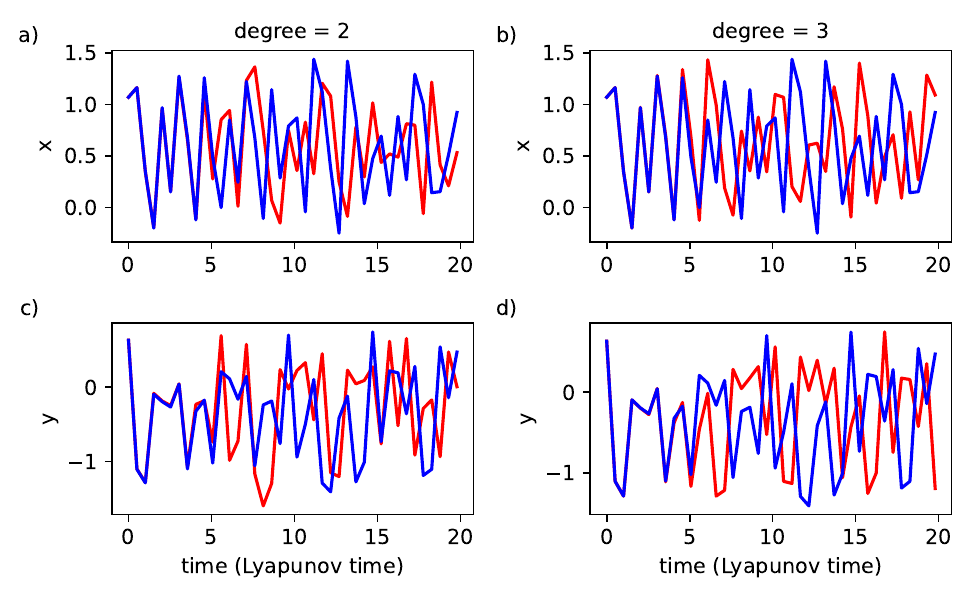}
    \caption{\textbf{Comparing the true and predicted Ikeda time series.} The dynamical evolution of the Ikeda variables as a function of the iteration number in units of the Lyapunov time ($\sim$2 iterates). Scale: $s=5$. Regularization: locally a) quadratic model with $\alpha=1\times 10^{-4}$ and b) cubic model with $\alpha=1\times 10^{-5}$.
    }
    \label{fig:time_series}
\end{figure*}

We use LB-NGRC models to forecast the Ikeda map starting from the first testing segment, as shown in Fig.~\ref{fig:time_series}.  We see that both the locally quadratic and cubic models have a forecasting horizon of about 5 Lyapunov times, beyond which the error begins because the system is chaotic. We conclude that the LB-NGRC model has a good short-term forecasting ability.  A forecasting horizon of 5 Lyapunov times is comparable to other good ML models.

\begin{figure}[htb!] 
    \centering
    \includegraphics[width=\columnwidth]{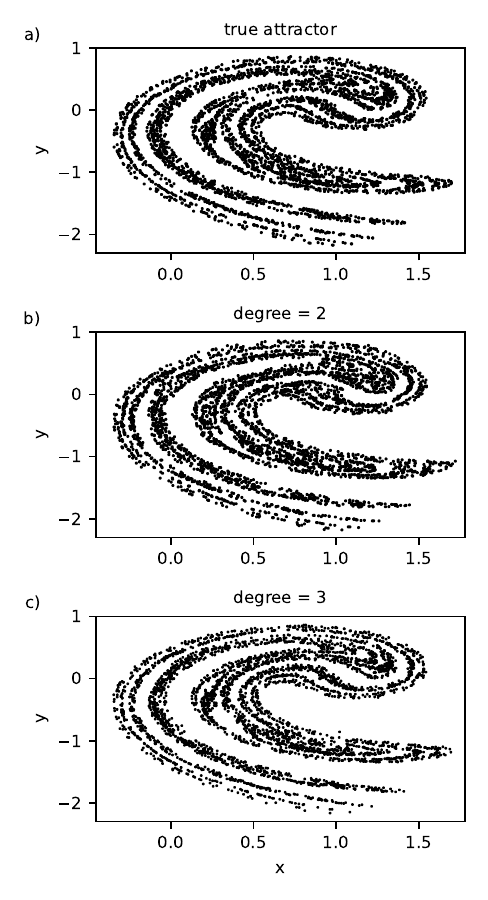}
    \caption{\textbf{Learning the climate of the Ikeda attractors.} Iteration of the Ikeda map and LB-NRGC models for 2,000 Lyapunov times ($\sim$4,000 iterates) for the a) true attractor, locally b) quadratic, and c) cubic LB-NGRC models. Scale: $s=5$. Regularization: locally a) quadratic model with $\alpha=1\times 10^{-4}$ and b) cubic model with $\alpha=1\times 10^{-5}$.
    }
    \label{fig:attractors}
\end{figure}

To assess the long-term prediction ability of LB-NGRC (the `climate'), Fig.~\ref{fig:attractors} shows the actual and predicted attractors for the locally quadratic and cubic models.  On a coarse scale, the attractors look similar with the basic structures reproduced.  However, for the locally quadratic model, there are some subtle differences in areas of the attractor that are slightly filled. These issues are largely resolved for the locally cubic model.

\section{Discussion}\label{sec:Discussion}

Our results demonstrate that low-order polynomial fits to data localized in phase space and smoothly blended allow for accurate forecasting of dynamical systems.  The local data sets are created by clustering the data using a hierarchal ball tree method, which is an unspervised method that does not require fine-tuning for our example problem.  For each cluster, we fit a low-order polynomial model to the data and blend the models for each region using radial basis functions.  

This general approach has already been established in the statistics community some time ago by, for example, Chaudhuri \textit{et al.}.\cite{Chaudhuri1994}  We identify that the ball-tree method is a good approach for clustering the data, which is well matched to our further improved model based on of radial basis functions for a convex combination model blending, and there is analysis of this in Appendix \ref{app:qual}.  
Therein we describe some advantages of LB-NGRC beyond the ability to handle systems where the global model would be to complicated such at the Ikeda map.  
Specifically, the error of a local regression model tends to improve depending on the domain size, the degree of the fit, and also considering the regularity of the fitting function in the local domain.   This allows leveraging of locality scale and partition, local model complexity, and spatial data density associated with data-set size with this new and efficient methodology.
In the future, we will explore using other clustering methods, including early stopping of the local branching based on the quality-of-fit for each parent ball,\cite{Chaudhuri1994} or random forests to create an ensemble model.\cite{Moradi2024}

Another method to simplify the models is to perform feature selection for each local model - that is, identify the monomials in each ball that contribute the most to predicting the data variance.  For example, forward regression orthogonal least squares is a highly successful approach for model reduction.  It is a greedy bottom-up algorithm that works efficiently and has higher numerical stability compared to other methods.\cite{Chen1989}

We mention that LB-NGRC has some similarity to other methods, such Gaussian-process models and support vector machines, but those only fit a constant-weighted Gaussian function.  Using local polynomials reduces the depth of the tree (\textit{i.e.}, reduces $s$) to obtain the desired model accuracy.  We also only evaluate the radial basis function at the ball centers, whereas the other  methods place a center at every training data point, which increases the computational cost of evaluating the model during deployment.

In the future, we will apply LB-NGRC to modeling ordinary differential equations, which will likely require using data from past steps.\cite{NGRC} Our approach can also be used for modeling time-delay or spatial-temporal dynamical systems, where the scalability of the ball tree method will likely be important.

\begin{acknowledgments}

EB has been supported by the NSF-NIH-CRCN, DARPA RSDN,
the ARO, and the ONR. 

\end{acknowledgments}

\section*{Conflict of Interest}
DJG and AP are commercializing reservoir computing.

\section*{Data Availability Statement}

The data that support the findings of this study are openly available in [repository name] at http://doi.org/[doi], reference number [reference number].

\appendix

\section{Quality of Fit of the LB-NGRC Model}\label{app:qual}

\begin{figure}[tb]
    \centering
    \includegraphics[width=\columnwidth]{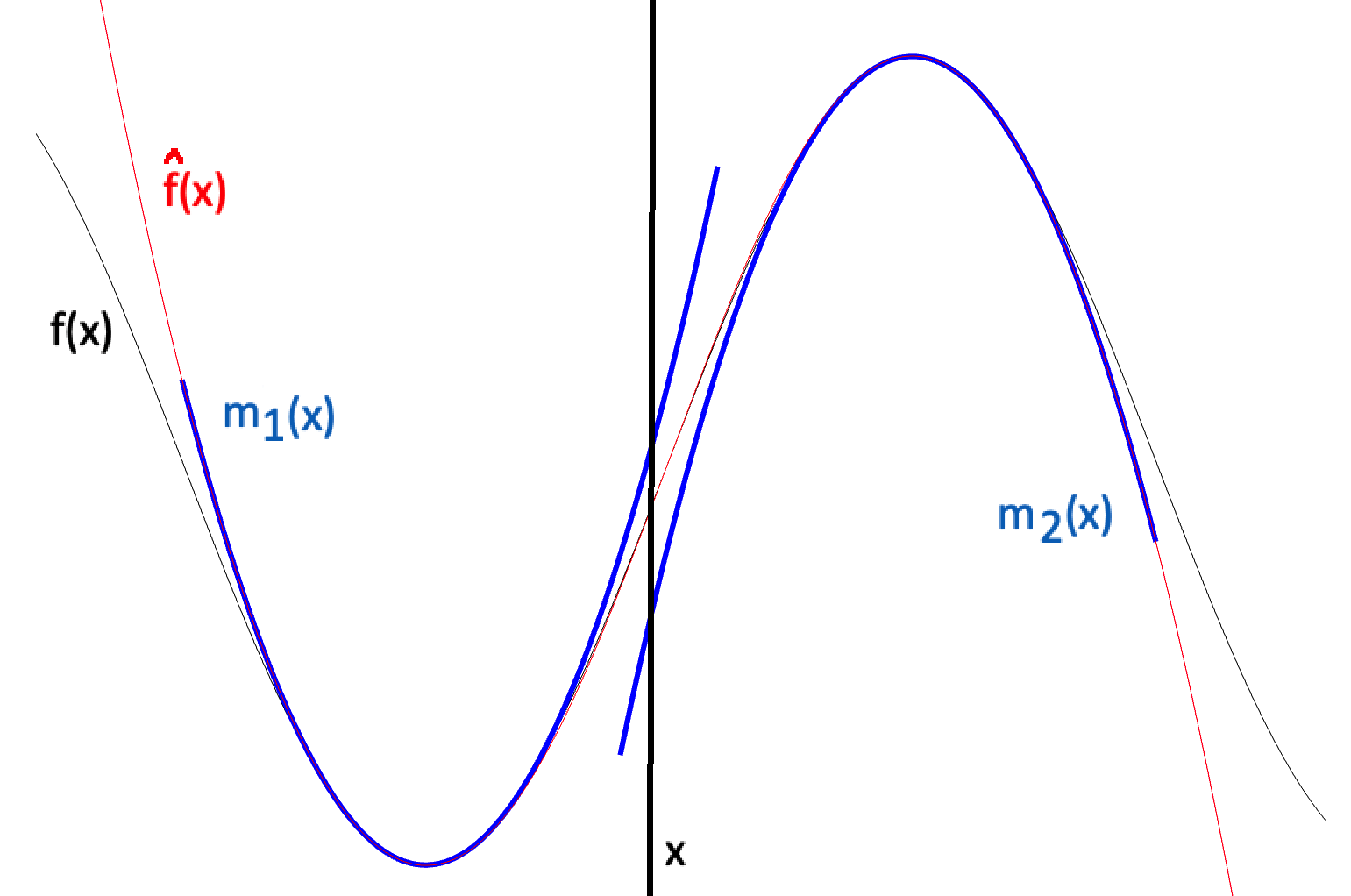}
    \caption{ A key feature of blended locality models Eqs.(\ref{eq:approxfit}), (\ref{eg:bended}) is that high-quality local models can be of relatively low order, but they degrade as the distance grows from their centers. The blending concept allows for a gradual ```hand-off" of the local functions to transition smoothly between zones of best performance.
    }
    \label{fig:simpleblend}
\end{figure}

Here, we present an analysis for the quality of fit of the locally blended model given in Eq.~(\ref{eq:blendedNGRC}), which is at the heart of this paper.   First, we present the idea in terms of two separate local models that we call $m_1(x)$ and $m_2(x)$ to a single scalar function $f(x)$, drawn for clarity in terms of the simplest setting of a one-dimensional domain and range.  See Fig.~\ref{fig:simpleblend}.

As imagined, $m_1(x)$ ($m_2(x)$) is a local model generated to produce a high-quality fit in the region on the left (right) side of the point $x$. Each model could be, for example, a Taylor series or some other kind of power series of basis functions designed to be expressive in each local region.  However, $x$ is somewhere in between where neither model is designed to be at its best.  In this middle region, either model is pretty good, but a blend between the two models may be the best choice.  By blend, we mean a convex combination.  

For simplicity, we assign a weight of 1/2 to each local model. Let the model fit for $x$ in the middle be given by
\begin{equation}\label{eq:approxfit}
\hat{f}(x)=\frac{1}{2} m_1(x)+ \frac{1}{2} m_2(x).
\end{equation}
The fit error is then 
\begin{eqnarray}\label{eq:err}
\textrm{err}(x)&:=&|f(x)-\hat{f}(x)| \\ &=&|\frac{1}{2}f(x)-\frac{1}{2}m_1(x)+\frac{1}{2}f(x)-\frac{1}{2}m_2(x)| \nonumber \\
&\leq & |\frac{1}{2}f(x)-\frac{1}{2}m_1(x)|+|\frac{1}{2}f(x)-\frac{1}{2}m_2(x)| \nonumber \\
&=& \frac{1}{2}\textrm{err}_1(x)+\frac{1}{2}\textrm{err}_2(x).
\end{eqnarray}
That is, the error is half of the error of the fit from using either one or the other model.

In a more general setting of a convex combination of $B$-local models $m_b(x)$, $b=1,2,..,B$, derived from a sum of weights $s_b(x)$ so that
\begin{equation}\label{eq:convex}
    s(x)=\sum_{i=b}^B s_b(x),
\end{equation}
then a blended function is defined by a convex combination given by
\begin{equation}\label{eg:bended}
    \hat{f}(x)=\frac{s_1(x)}{s(x)}m_1(x)+...+\frac{s_B(x)}{s(x)}m_B(x).
\end{equation}
The error can be bounded by derivation analogous to the simpler case of Eq.~(\ref{eq:err}) and is found to be
\begin{eqnarray}
     \textrm{err}(x)&\leq &|\frac{s_1(x)}{s(x)}||f(x)-m_1(x)|+...+|\frac{s_B(x)}{s(x)}||f(x)-m_B(x)| \nonumber \\
     &=& \sum_{i=b}^B |\frac{s_b(x)}{s(x)}||f(x)-m_b(x)| \nonumber \\
    & \leq& \frac{1}{{ S}}\sum_{i=b}^B |s_b(x)|f(x)-m_b(x)|, \mbox{ where }{S}\geq |s(x)|\nonumber \\
    &\leq & R~M, 
\end{eqnarray}
in terms of a bounding value ${ S}$.  Let, $R=1/S$ and
\begin{equation}
    M=\max_{b\in \{1,2,..,B\}} \sup_{x\in {\cal B}_b} |f(x)-m_b(x)|,
\end{equation}
for ${\cal B}_b$ a bounded domain over which $m_b(x)$ is defined. Here, the domain is generally one of the balls of the ball tree construction.  The quantities $R$ and $M$ are bounded and have a naturally controlled size by relying on the concept inherent to the ball tree partitioning method.

In each ball ${\cal B}_b$, when the ball is relatively small, the idea is that a low-degree local NGRC model $m_b(x)$ works progressively better even for low-degree models as long as the data set size within a ball stays approximately the same. This is the same concept as a Taylor-series expansion of an analytic function, where the quality of fit depends on distance from the center and the degree of the series. Similarly, a standard analysis \cite{***} for a Lagrange polynomial $P_r(x)$ of degree $r$ fit through $r+1$ points in a extent of data $\{(x_1,f(x_1),(x_2,f(x_2),...,(x_{r+1},f(x_{r+1})\}$ for $\{x_1,x_2,...,x_r\}\in {\cal B}$, there is point $c\in {\cal B}$ such that
\begin{equation}\label{scaling}
   | P_r(x)-f(x)|=
   \left|\frac{f^{(r+1)}(c) \Pi_{i=1}^r (x-x_i)}{(r+1)!} \right|\leq K \frac{|x-x_b|^r}{(r+1)!}.
\end{equation}
Here, $x_b$ is a central point in ${\cal B}$, and let $K=\max_{x\in {\cal B}} |f^{(r+1)}(x)|/(r+1)!$ is a standard statement on error bounding by interpolating polynomials in one dimension; comparable statements in terms of derivatives exist for the multivariate scenario even though it is more complicated to write.  
With  $M_b:=\sup_{x\in {\cal B}_b} |f(x)-m_b(x)|$, then practically 
$M_b\sim K |x-x_b|^r/(r+1)!$ from Eq.~(\ref{scaling}).

The point of the above discussion is that $M$ is well controlled for each ball domain ${\cal B}_b$, at least if the balls decrease in diameter less than one, and the regularity of the derivatives bounding $K$ do not outpace the decreasing of $ |x-x_*|^r/(r+1)!$ with increasing $r$. This bound becomes progressively better with small balls, resulting from large data sets that result in high data density even in small balls, which, inn turn, allow for lower-degree polynomial fits.  This is the strength of local models.
Likewise and similarly, the error of a local regression model tends to improve depending on the domain size, the degree of the fit, and also considering the regularity of the fitting function in the local domain.  

Specifically, in Eqs.~(\ref{eq:blendedNGRC})-(\ref{eq:bend2}), each weight function is defined as a radial basis function
\begin{equation}
    s_b(x)=\textrm{RBF}_b(\tilde{x_b},\sigma_b),
\end{equation}
which will be in the analysis below. 
It is easy to see that Eq.~(\ref{eg:bended}) represents a convex sum for each $x$, since 
\begin{equation}\label{eg:bended2}
   \frac{s_1(x)}{s(x)}+...+\frac{s_B(x)}{s(x)}=1,
\end{equation}
and considering Eq.~(\ref{eq:convex}).
What is notable is that using radial basis functions, with standard deviations $\sigma_b$ that are generally a fraction $p>1$ of the ball radius size,  it effectively (almost) results an averaging across those models developed in the balls nearest to a point $x$ where the bended model $\hat{f}(x)$ is to be evaluated.  In fact, it is straightforwardly obvious that since exponentials decay so fast, that a set distance from a ball center can be chosen so that the RBF, $\sigma_b(x)$, will be sufficiently small that the blended sum model, Eq.~(\ref{eg:bended}).  In fact, this brings up an efficiency opportunity to simply skip those $m_b(x)$ locality models in the summation of $\hat{f}(x)$ corresponding to the less nearby balls (most of them) whenever $\sigma_b(x)<\delta$ for some small threshold $\delta>0$.  However, the quality of the approximation by such a reduced blended model using fewer terms by only the nearby terms, $k=1,...,K_\delta$, $K_\delta<B$ will be better if we recalibrate the sum of $s(x)$ in Eq.~(\ref{eq:convex}) to include only the those terms corresponding to the nearby balls.

\nocite{*}
\bibliography{NGRCwithAttention}

\end{document}